# Porous carbon nanowire array for highly sensitive, biocompatible, reproducible surface-enhanced Raman spectroscopy

Nan Chen[1,2,†], Ting-Hui Xiao[1,†], Zhenyi Luo[1], Yasutaka Kitahama[1], Kotaro Hiramatsu[1,3,4], Naoki Kishimoto[5], Tamitake Itoh[6], Zhenzhou Cheng[1,7] & Keisuke Goda[1,8,9*]

1   Department of Chemistry, The University of Tokyo, Tokyo 113-0033, Japan
2   School of Chemistry and Chemical Engineering, Beijing Institute of Technology, Beijing 100081, P. R. China
3   Research Centre for Spectrochemistry, The University of Tokyo, Tokyo 113-0033, Japan
4   PRESTO, Japan Science and Technology Agency, Saitama 332-0012, Japan
5   Department of Chemistry, Tohoku University, Sendai 980-8578, Japan
6   Health Research Institute, National Institute of Advanced Industrial Science and Technology, Takamatsu 761-0395, Japan
7   School of Precision Instruments and Opto-electronics Engineering, Tianjin University, Tianjin 300072, China
8   Institute of Technological Sciences, Wuhan University, Hubei 430072, P. R. China
9   Department of Bioengineering, University of California, Los Angeles, California 90095, USA

*E-mail: goda@chem.s.u-tokyo.ac.jp

**Surface-enhanced Raman spectroscopy (SERS) is a powerful tool for vibrational spectroscopy as it provides several orders of magnitude higher sensitivity than inherently weak spontaneous Raman scattering by exciting localized surface plasmon resonance (LSPR) on metal substrates[1-4]. However, SERS is not very reliable[5-7], especially for use in life sciences, since it sacrifices reproducibility and biocompatibility due to its strong dependence on "hot spots"[8-10] and large photothermal heat generation[11,12]. Here we report a metal-free (i.e., LSPR-free), topologically tailored nanostructure composed of porous carbon nanowires in an array as a SERS substrate that addresses the decades-old problem. Specifically, it offers not only high signal enhancement due to its strong broadband charge-transfer resonance, but also extraordinarily high reproducibility or substrate-to-substrate, spot-to-spot, sample-to-sample, and time-to-time consistency in SERS spectrum due to the absence of hot spots and high compatibility to biomolecules due to its fluorescence quenching capability. These excellent properties make SERS suitable for practical use in diverse biomedical applications.**

Since its discovery in the 1970s, SERS has gained popularity as an analytical method for its extraoridinarily high signal enhancement[13-15]. Unfortunately, the benefit of SERS is compromised by its extremely poor reproducibility and biocompatibility due to its dependence on hot spots for high enhancement factors from aggregates of metal nanoparticles[16,17] or engineered metal nanostructures[4,9,10] and the generation of large photothermal heat on the metal surface that causes detrimental effects to biomolecules (e.g., heat-induced protein denaturation)[11,12]. In recent years, non-metallic materials such as silicon and germanium nanostructures[18-21], two-dimensional materials (e.g., graphene, $MoS_2$, and h-BN)[22-25], and semiconducting metal oxides[26-27] have been proposed as alternative SERS substrates to circumvent the fundamental problem of traditional SERS. Different from the metal substrates that mainly rely on LSPR, their signal enhancement originates from structural resonance[19-21] or charge transfer resonance[25,26] and has been demonstrated with decent enhancement factors of up to five orders of magnitude[19-21,25,26]. However, the poor reproducibility remains a major challenge while these substrates partly address the biocompatibility issue. This is due to their inherent photocatalytic activity and the toxicity of their substrate material to biomolecules[28-30].

In this Letter, we take a radically different approach from the metal, semiconducting, and dielectric substrates and demonstrate that a metal-free (i.e., LSPR-free), topologically tailored nanostructure composed of a two-dimensional array of porous carbon nanowires is an effective material as a substrate for highly sensitive, biocompatible, and reproducible SERS (Supplementary Table 1). Specifically, the porous carbon nanowire array (PCNA) substrate provides not only high signal enhancement (~$10^6$) due to its strong broadband charge-transfer resonance for large chemical enhancement (as opposed to electromagnetic enhancement in traditional SERS), but also extraordinarily high reproducibility or substrate-to-substrate, spot-to-spot, sample-to-sample, and time-to-time consistency in SERS spectrum (which is not possible with traditional SERS) due to the absence of hot spots and high compatibility to biomolecules due to its fluorescence quenching capability. We experimentally demonstrate these excellent properties with various molecules such as rhodamine 6G (R6G), β-lactoglobulin, and glucose. To the best of our knowledge, our PCNA substrate offers the highest reproducibility and biocompatibility to date, holding great promise for reliable SERS in practical use, especially in areas of analytical chemistry, pharmaceutical science, food science, forensic science, and pathology where inconsistent or non-reproducible SERS spectra have been problematic.

The synthesis and material properties of the PCNA are described as follows and shown in Fig. 1a through Fig. 1d (see "experimental materials", "growth of the PCNA", and "fabrication of the PCNA substrate" in Supplementary Information for details). First, a polypyrrole (PPy) nanowire array (PNA) is prepared through a conventional template-assisted electropolymerization method[31] with an anodized aluminum oxide (AAO) template (Supplementary Fig. 1). Second, a working electrode is interchanged with a counter electrode. Then, the PNA undergoes an electrical degradation process in a high-temperature dimethyl sulfoxide (DMSO) solution containing sulfur clusters under an oppositely applied voltage to form a large number of nanopores in each PPy nanowire, which transforms into a porous polypyrrole nanowire array (PPNA) and effectively increases the specific surface area (SSA) and roughness. Finally, a carbonization process at an elevated temperature is applied to the PPNA to generate a SERS-active PCNA (Supplementary Figs. 2 and 3, see "thermogravimetric analysis of the PNA" in Supplementary Information for details). Each porous nanowire of the PCNA has an average diameter of 140 nm and an average length of 15 µm after the carbonization (Fig. 1a and Supplementary Fig. 4, see "optical properties of the PCNA" in Supplementary Information for details), which is in agreement with those of the AAO template. An enlarged scanning electron microscope (SEM) image of the PCNA reveals that numerous holes with an average diameter of about 50 nm are distributed on the PCNA, resulting in fractal nanostructures with a high SSA (Supplementary Figs. 5 and 6, see "SEM imaging of PNA, CNA, and PCNA substrates" and "nitrogen adsorption-desorption isotherm measurements of the PNA and PCNA substrates" in Supplementary Information for details). In addition, as shown in Fig. 1b, results from our Raman spectroscopy of the PPNA and PCNA indicate that all the characteristic Raman peaks of the PPNA at 870, 930, 1050, and 1246 cm$^{-1}$ disappeared from the PCNA after the carbonization, as expected. As shown in Fig. 1c, results from our current-voltage (*I-V*) curve measurements of the PPNA and PCNA indicate that the porous carbon nanomaterial performs as a semiconductor since the carbonization treatment significantly increases its conductivity (Supplementary Figs. 7 and 8, see "methods for characterizing PPNA and PCNA substrates" and "X-ray diffraction measurements of the PCNA" in Supplementary Information for details). Moreover, results from our energy-dispersive X-ray spectroscopy (EDS) of the PPNA and PCNA indicate that the composition ratio of carbon significantly increased after the carbonization process as shown in Fig. 1d.

To demonstrate the high adsorptivity and reproducibility of the PCNA as a SERS substrate, we performed SERS of R6G (a highly fluorescent rhodamine family dye) on the PCNA substrate. For comparison, we obtained Raman spectra of R6G molecules adsorbed on silicon, PNA, carbon nanowire array (CNA), and PCNA substrates under the same conditions (a 10-µM deionized water solution of R6G molecules, an integration time of 30 s, continuous-wave laser illumination with 1 mW at 785 nm), as shown in Fig. 2a (Supplementary Fig. 9 and Supplementary Table 2, see "Raman peak assignments of R6G, β-lactoglobulin, glucose, and DMSO" in Supplementary Information for details). In comparison to the other substrates, the PCNA exhibits the highest Raman signal intensity, indicating its high adsorptivity (i.e., large surface-to-volume ratio), which originates from its porous nanowire morphology and carbon-based material. Moreover, to evaluate the sensitivity of the PCNA substrate, we obtained SERS spectra of R6G molecules at reduced concentrations on the PCNA substrate (Fig. 2b). It is evident from the figure that the detection limit of the PCNA substrate for R6G molecules is about 10 nM. It is important to note that the use of carbon quenched the fluorescent property of R6G molecules, indicating the ability of the PCNA substrate to analyze biomolecules which are mostly fluorescent (Supplementary Fig. 10). Similar SERS measurements were performed on DMSO to verify the high sensitivity of the PCNA substrate (Supplementary Fig. 11, see "Raman spectroscopy of DMSO on the silicon and PCNA substrates" in Supplementary Information for details). As shown in Fig. 2c, the PCNA substrate shows a monotonically increasing relation between the R6G concentration and signal intensity at 1185, 1309, 1361, 1507, and 1650 cm$^{-1}$. Furthermore, in order to assess the substrate-to-substrate consistency of the PCNA substrate, we carried out a measurement reproducibility test on 20 different PCNA substrates. As shown in Fig. 2d, the differences in the relative intensities of the Raman peaks at 1185, 1309, 1361, 1507, and 1650 cm$^{-1}$ between all of the substrates are with a standard deviation of 5.7%, indicating the high reliability of the PCNA substrate. It is important to mention that the Raman intensity of the PCNA substrate (Fig. 1b) does not interfere with the enhanced Raman spectra of probed molecules since it is much weaker than them and its characteristic peaks are very broad and submerged by them (Supplementary Fig. 12).

Next, to demonstrate the highly sensitive detection of biomolecules, we conducted SERS of β-lactoglobulin (the major whey protein of cow and sheep's milk) on the PCNA substrate. With an integration time of 300 s and an excitation illumination of 2 mW at 785 nm, we measured a spontaneous Raman spectrum of a β-lactoglobulin powder with a mass fraction (Mf) of 100% on a bare silicon substrate as ground truth (Fig. 3a, Supplementary Fig. 9, and Supplementary Table 2, see "Raman peak assignments of R6G, β-lactoglobulin, glucose, and DMSO" in Supplementary Information for details). Then, by decreasing the integration time to 1 s and the Mf to 0.4%, we obtained the Raman spectrum of β-lactoglobulin molecules with a similar Raman signal intensity on the PCNA substrate. As the ratio of the number of molecules of a β-lactoglobulin solution to that of the β-lactoglobulin powder in the probed volume is approximately equal to their Mf ratio (Supplementary Table 3), we used the Mf ratio to calculate the SERS enhancement, based on which the enhancement factor was found to be (300 s / 1 s) × (100% / 0.4%) × (60000 / 2000) = ~$10^6$ for β-lactoglobulin. It is important to note that the characteristic Raman peaks of β-lactoglobulin molecules on the PCNA substrate are well distinguishable and consistent with those of the ground truth, indicating the high biocompatibility of the PCNA substrate. For comparison, we also acquired the Raman spectra of β-lactoglobulin molecules on a silicon substrate and a commercial metal SERS substrate (sliver-gold hybrid substrate, SERSitive Co.) under the same conditions, as shown in Fig. 3a and Supplementary Fig. 13. On the silicon substrate, no characteristic Raman peaks were visible due to the absence of the SERS enhancement whereas on the metal substrate, the Raman spectrum was enhanced, but with 10 times weaker than on the PCNA substrate. Also, the characteristic Raman peaks of β-lactoglobulin molecules on the metal substrate do not agree with those of the ground truth obtained from the silicon substrate (Fig. 3a). Finally, to assess the spot-to-spot consistency of the PCNA substrate in Raman signal intensity, we conducted SERS mapping of β-lactoglobulin on the PCNA substrate at two characteristic Raman peaks of the molecule (999 cm$^{-1}$ and 1447 cm$^{-1}$) on both large and small scales (Fig. 3b). As shown in Fig. 3c, the PCNA substrate has a coefficient of variation (CV) of less than 7.8% on average, which is much smaller than that of the conventional metal substrate (Supplementary Fig. 14). These results firmly demonstrated that the PCNA substrate would be an excellent platform for reliable trace detection of proteins which are typically vulnerable to heat and are difficult to probe on metal substrates.

Finally, to demonstrate the sample-to-sample consistency of the PCNA substrate in the enhancement factor, we conducted SERS of glucose (a well-known biomarker for detecting diabetes[32], which has been a challenging molecule for undistorted SERS detection. Fig. 3d shows the Raman spectra of a deionized water solution of glucose molecules on silicon and PCNA substrates under light excitation at 785 nm, with the former used as a ground truth (Supplementary Fig. 9 and Supplementary Table 2, see "Raman peak assignments of R6G, β-lactoglobulin, glucose, and DMSO" in Supplementary Information for details). On the PCNA substrate, all the characteristic Raman peaks of glucose molecules were clearly identified and distinguished even at a low Mf of 0.1%, corresponding to a concentration of 5.6 mM (which agrees with the typical glucose concentration of 3-7 mM in the blood of healthy people as opposed to 10-20 mM in the blood of diabetes patients[33]. Similar to that of β-lactoglobulin, we used the Mf ratio of glucose to calculate the SERS enhancement (Supplementary Table 3), based on which the comparison of the Raman spectra on the silicon and PCNA substrates yields an enhancement factor of (300 s / 1 s) × (100% / 0.1%) × (1600 / 500) = ~$10^6$, which is consistent with the enhancement factor of the β-lactoglobulin measurements above. Importantly, even with this high signal enhancement, the locations of the measured characteristic Raman peaks of glucose molecules agree well with those of the ground truth (while each Raman peak has a different chemical enhancement), which is attributed to the high biocompatibility and photothermal stability of the PCNA substrate. Furthermore, to show the time-to-time consistency of the PCNA substrate, we performed SERS measurements of glucose under the same conditions every hour. As shown in Fig. 3e, the Raman spectrum of glucose molecules is temporally stable with negligible deterioration (e.g., no oxidization), further demonstrating the high reliability of the PCNA substrate (Supplementary Figs. 15 and 16).

The SERS signal enhancement of the PCNA is mainly attributed to the chemical mechanism (CM) for the following reasons. First, carbon provides a high charge-transfer efficiency[15], which greatly increases its Raman-scattering cross-section. Fig. 4a shows an energy level diagram (obtained by our theoretical analysis based on density functional theory with Gaussian16[34]) that indicates the molecular orbitals of a test molecule (R6G) on a carbon sheet that approximates the surface of the PCNA (Supplementary Fig. 17, see "theoretical analysis of the chemical mechanism" in Supplementary Information for details). The diagram also indicates two charge transfer pathways from the highest occupied molecular orbital (HOMO) to the PCNA-R6G hybrid states, enabling the R6G molecule on the surface of the PCNA to resonantly excited at the wavelengths of 785 nm (1.58 eV) and 532 nm (2.33 eV), whereas the R6G molecule alone cannot be resonantly excited at 785 nm due to the absence of excited states between the HOMO and the lowest unoccupied molecular orbital (LUMO) and can be resonantly excited at 532 nm, but with a very strong fluorescence background that obscures the Raman spectrum. These theoretically predicted resonant Raman scattering effects, which account for the extraordinary SERS enhancements at the excitation wavelengths of 532 nm and 785 nm, are in good agreement with our experimental results as shown in Fig. 4b. Second, H, N, and S atoms in the hydroxyl group that are not completely removed after the carbonization (Fig. 1d) further promote the charge transfer between the substrate and molecules[15], contributing to the SERS enhancement. Third, based on our finite element method simulation (Fig. 4c), the largest electric field enhancement localized at the lateral edges of each PCNA nanowire is only about 2, corresponding to a small average enhancement factor of ~1.8 (Supplementary Fig. 18). Also, the small difference between the electric field magnitude distributions at the excitation wavelengths of 532 nm and 785 nm shows the absence of strong structural resonance, indicating a small electromagnetic contribution to the overall SERS enhancement factor.

To further verify this theory, we performed the following experiments. First, we measured the absorption spectrum of the PCNA substrate with and without β-lactoglobulin on it as shown in Fig. 4d. The broad bandwidth of the absorption spectrum excludes the electromagnetic mechanism (EM) which normally possesses a narrow bandwidth due to its structural resonance. Fig. 4d also shows the charge-transfer band obtained by taking the difference of the two absorption spectra (Supplementary Fig. 19). The broad charge-transfer band that covers the excitation wavelength (785 nm) indicates an electronic band essential for the efficient charge-transfer resonance. Second, we measured the Raman spectrum of the same molecule up to the high Raman shift region (up to 3200

cm$^{-1}$) which covers overtones and combination bands of the molecule as shown in Fig. 4e. No peaks of overtones and combination bands were observed in the region. As overtones and combination bands are nonlinear optical effects whose magnitudes are determined by the electric field magnitude and normally evident in EM-based SERS, the absence of their peaks in this region excludes the EM, further verifying the CM as the dominant effect of the Raman enhancement. Third, we measured the Raman spectrum of β-lactoglobulin at different excitation wavelengths (785 nm and 532 nm). As shown in Fig. 4f, enhancement factors of ~$10^6$ and ~$10^5$ at an excitation wavelength of 785 nm and 532 nm, respectively, were observed, verifying the broadband CM enhancement of the PCNA substrate.

By virtue of the combined merits of the high reproducibility and excellent biocompatibility of the PCNA substrate, the current enhancement factor of $10^6$ is sufficient for practical use for Raman spectroscopy of biomolecules at low concentrations as it is comparable to or even higher than the enhancement factor of commercial metal SERS substrates. A higher SERS enhancement factor can presumably be achieved by optimizing the composition (e.g., types and doping levels of dopants) and structure (e.g., porous nanowire morphology) of the PCNA substrate, employing an excitation laser whose wavelength is optimal for the CM as the enhancement factor depends on the probed molecule (e.g., via a wavelength-tunable laser), or improving the theory of the CM for a better understanding of its underlying principles. With these improvements, PCNA-based SERS is expected to enable a wider range of biomedical applications such as quantitative analysis of chemical bonds in proteins, accurate evaluation of glucose in blood for diabetes detection, and trace detection of toxic substances in food and water.

**Data availabililty**

All data supporting the findings of this study are included in the article and its Supplementary Information, and are also available from the authors upon reasonable request.

**Code availabililty**

All codes used for analysis of this study are available from the authors upon reasonable request.

**Acknowledgements** This research was supported by JSPS Core-to-Core Program, JSPS KAKENHI (JP18K13798), White Rock Foundation, University of Tokyo GAP Fund, National Natural Science Foundation of China (21671020), Beijing Natural Science Foundation (2172049), KISTEC, Nakatani Foundation, and Ogasawara Foundation for the Promotion of Science and Engineering.

**Author Contributions** N. C. and T. H. X. are co-first authors and contributed equally to this work. N. C. and T. H. X. conducted the experiments and theoretical calculations. N. C., T. H. X., Y. K., K. H., Z. C. and K. G. interpreted the data. Z. C. led the work in the lab. K. G. supervised the research team. N. C., T. H. X., Z. C. and K. G. wrote the manuscript. All authors discussed the results and participated in revising the manuscript.

**Author Information** Reprints and permissions information is available at www.nature.com/reprints. The authors declare no competing financial interests. Correspondence should be addressed to K. G. (goda@chem.s.u-tokyo.ac.jp).

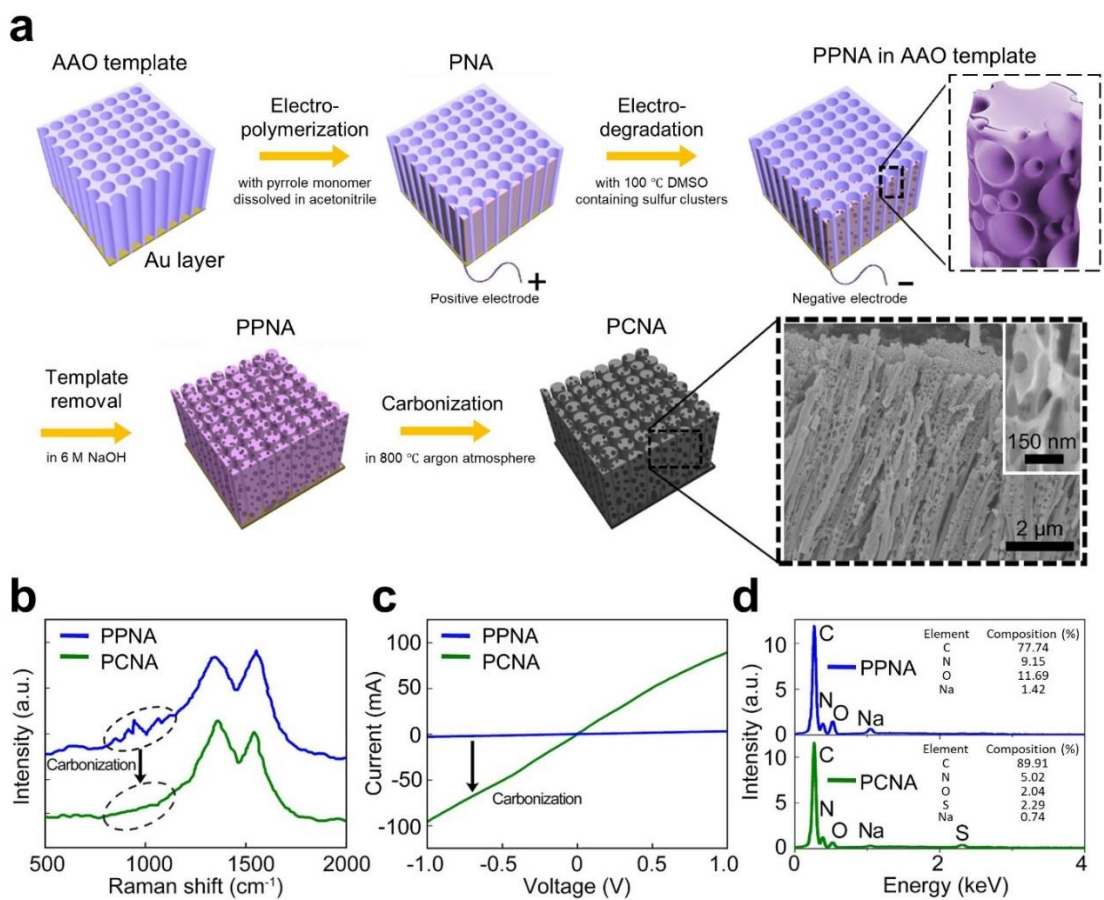

**Fig. 1 | Synthesis and characterization of the porous carbon nanowire array (PCNA). a**, Steps for synthesizing the PCNA. The bottom right inset shows an SEM image of the PCNA and an enlarged SEM image of a single porous carbon nanowire. **b**, Raman spectra of the PPNA and PCNA. After the carbonization process, all the characteristic Raman peaks of the PNA at 870, 930, 1050, and 1246 cm$^{-1}$ disappeared as evident in the PCNA spectrum. **c**, *I-V* curve measurements of the PPNA and PCNA. In the measurements, the length and contact area of the porous PPy nanowires and porous carbon nanowires are about 15 μm and 0.5 mm$^2$, respectively. The conductivity of the substrate was significantly increased after the carbonization process, indicating the semiconducting property of the PCNA. **d**, EDS spectra of the PPNA and PCNA. The composition ratio of carbon was significantly increased after the carbonization process.

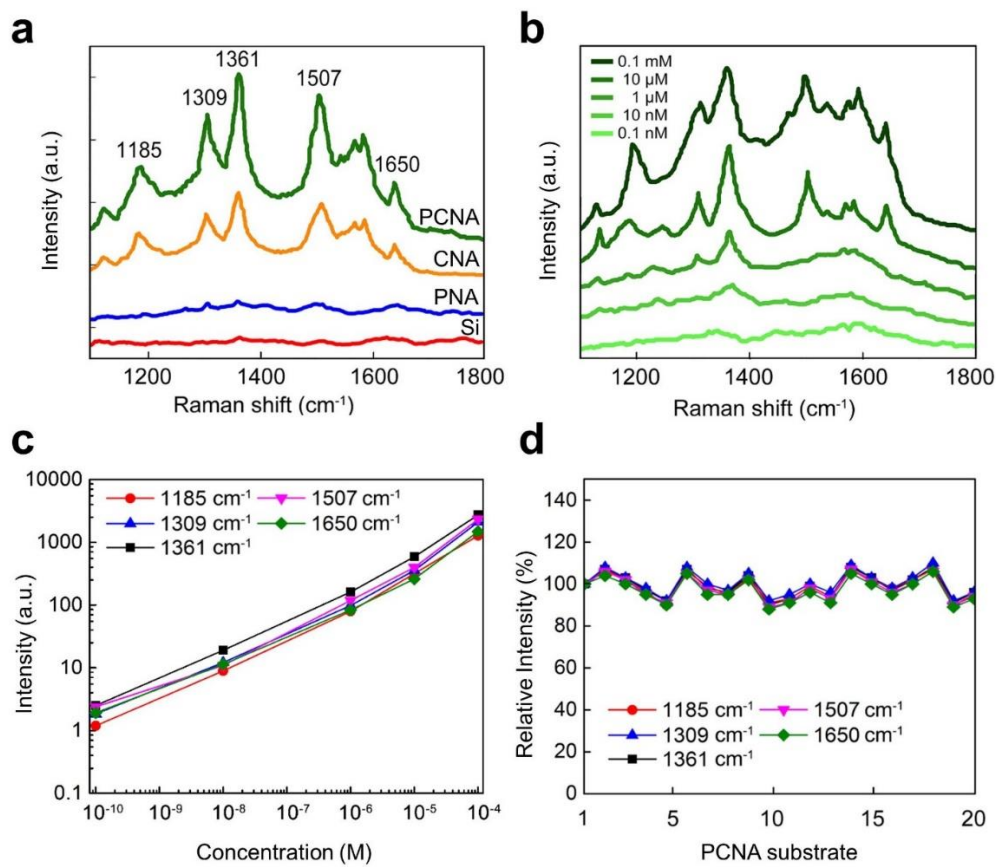

**Fig. 2 | SERS of R6G on the PCNA substrate. a**, Measured Raman spectra of R6G molecules at a concentration of 10 μM on the silicon (red), PNA (blue), CNA (orange), and PCNA (green) substrates for an integration time of 30 s with an excitation power of 1 mW at 785 nm after cleaning the substrates. **b**, Measured Raman spectra of R6G molecules at different concentrations adsorbed on the PCNA substrate for an integration time of 30 s with an excitation power of 1 mW at 785 nm. **c**, Intensities of the Raman peaks at different concentrations at 1185, 1309, 1361, 1507, and 1650 cm$^{-1}$. The detection limit of the PCNA substrate for R6G molecules is about 10 nM. **d**, SERS reproducibility measurements of R6G molecules on different PCNA substrates. The differences in the relative intensities of the Raman peaks at 1185, 1309, 1361, 1507, and 1650 cm$^{-1}$ between 20 different substrates are within a standard deviation of 5.7%.

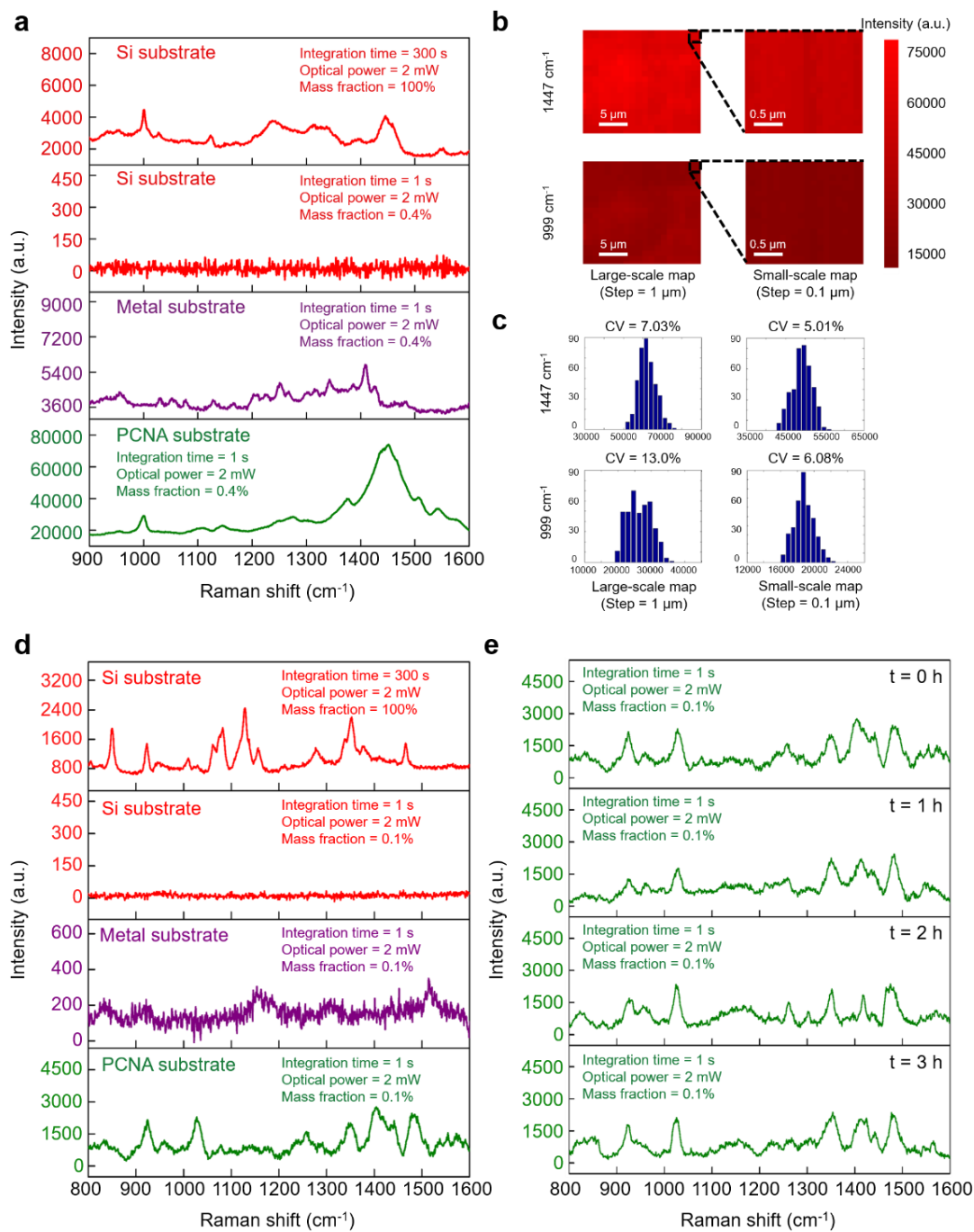

**Fig. 3 | SERS of biomolecules (β-lactoglobulin and glucose) on the PCNA substrate. a**, Measured Raman spectra of β-lactoglobulin molecules on the silicon, PCNA, and commercial metal substrates. The Raman signal intensity of β-lactoglobulin molecules on the PCNA substrate is 10 times higher than on the metal substrate. **b**, SERS maps of β-lactoglobulin on the PCNA substrate, showing high surface homogeneity in enhancement factor at two characteristic Raman shifts of the molecule on both large and small scales with a step size of 1 µm and 0.1 µm, respectively. **c**, Histograms of the enhancement factors on the large and small scales. **d**, Measured Raman spectra of glucose molecules on the silicon and PCNA substrates. **e**, Time-to-time consistency of the PCNA substrate in the Raman spectrum of glucose. Small hour-to-hour fluctuations in the Raman spectrum indicate high reproducibility and biocompatibility.

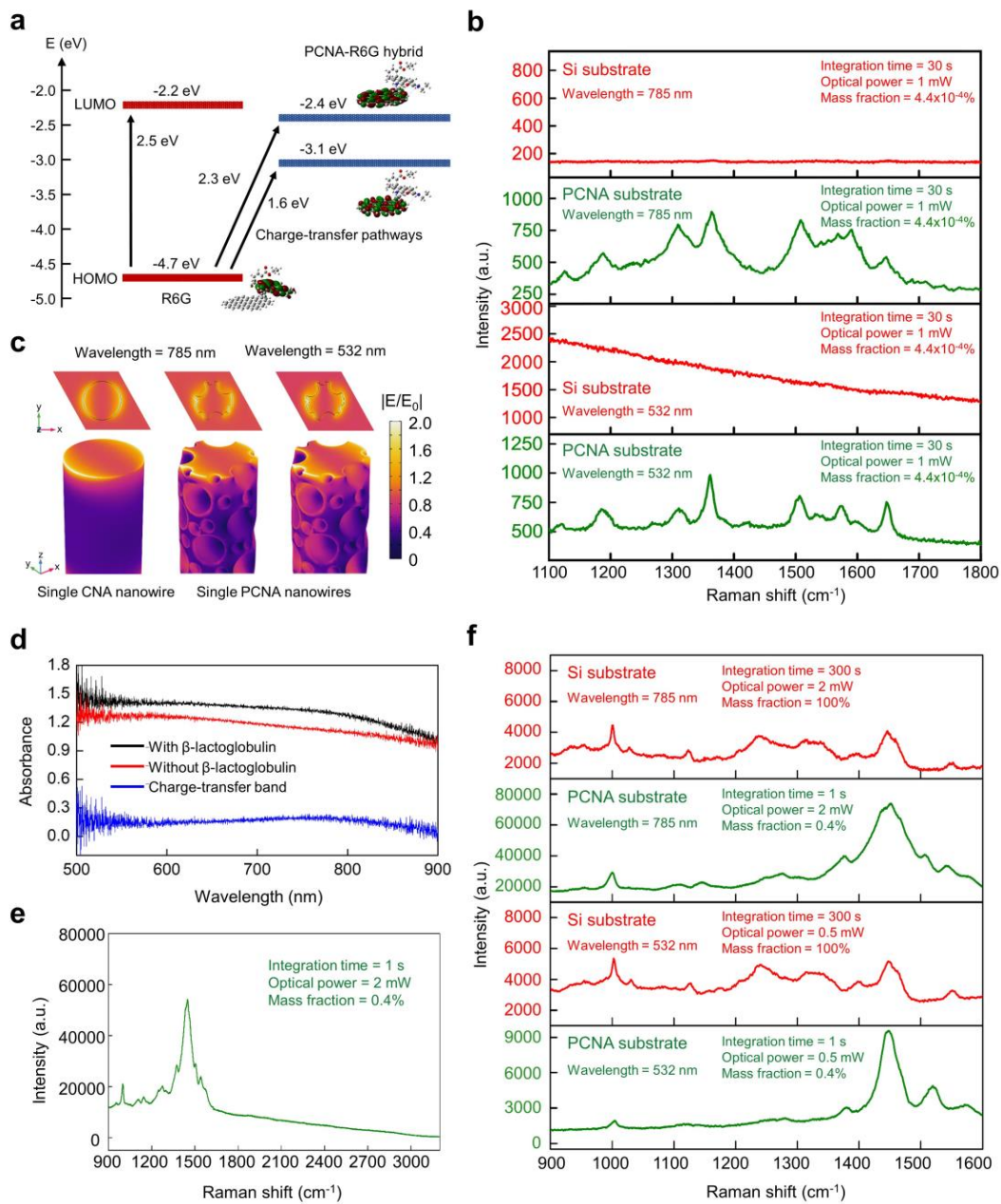

**Fig. 4 | Theory and experimental verification of the CM as the dominant effect on the PCNA substrate. a,** Energy level diagram (obtained by our theoretical analysis based on density functional theory with Gaussian16) that shows the molecular orbitals of a test molecule (R6G) on a carbon sheet that approximates the surface of the PCNA. The diagram shows two charge transfer pathways from the HOMO to the PCNA-R6G hybrid states, enabling the R6G molecule on the surface of the PCNA to resonantly excited at the wavelengths of 785 nm (1.58 eV) and 532 nm (2.33 eV), whereas the R6G molecule alone cannot be resonantly excited at these wavelengths due to the absence of the excited states. **b,** Raman spectra of R6G on the silicon and PCNA substrates at excitation wavelengths of 532 nm and 785 nm. **c,** Comparison in electric field magnitude distribution between the single CNA nanowire and the single PCNA nanowire to visualize the small contribution of the EM. **d,** Absorption spectrum of the PCNA substrate with and without β-lactoglobulin on it and the charge transfer band obtained by taking the difference of the two absorption spectra. **e,** Raman spectrum of β-lactoglobulin up to the high Raman shift region, showing no appreciable peaks of overtones and combination bands. **f,** Raman spectra of β-lactoglobulin at different excitation wavelengths of 532 nm and 785 nm.